# Intricate modulation of interlayer coupling at GO/MoSe$_2$ interface: application in time-dependent optics and device transport

Tuhin Kumar Maji[1], Kumar Vaibhav[2], Samir Kumar Pal[1], Kausik Majumdar[3], K. V. Adarsh[4] and Debjani Karmakar[5,*]

[1] *Department of Chemical Biological and Macromolecular Sciences, S.N. Bose National Centre for Basics Sciences, Salt Lake, Sector 3, Kolkata 700106, India*

[2] *Computer Division, Bhabha Atomic Research Centre, Trombay 400085, India*

[3] *Department of Electrical Communication Engineering, Indian Institute of Science, Bangalore 560012, India*

[4] *Department of Physics, Indian Institute of Science Education and Research, Bhopal 462066, India*

[5] *Technical Physics Division, Bhabha Atomic Research Centre, Trombay 400085, India*

*Corresponding author:
Debjani Karmakar:  karmakar.debjani@gmail.com

## *Abstract*

In GO/MoSe$_2$ semiconductor heterostructure, we have demonstrated a subtle control on the doping dynamics by modulating interlayer coupling through the combination of strain-reducing relative rotation of the constituting layers and variation of ligand type and concentration. By first-principles calculations incorporating spin-orbital coupling, we have investigated the impact of variable interlayer coupling in introducing non-collinear magnetic behaviour in the heterostructure. The outcome of varying carrier type and their respective concentrations are investigated by static as well as time dependent density functional calculations, which indicates presence of optical anisotropy and time-dependent optical phenomena like exciton quenching and band-gap renormalization. Performance of such heterostructures as channel material in devices with top and edge metal contacts is analyzed. Our self-consistent quantum transport calculations have evinced that the nature of interface-induced variation in doping is extrapolated for devices only in the case of top contacts. The edge contact, although exhibits a better transmission, are inefficient for sensing the ligand-induced doping modulation introduced via vertical inter-layer charge transfer.

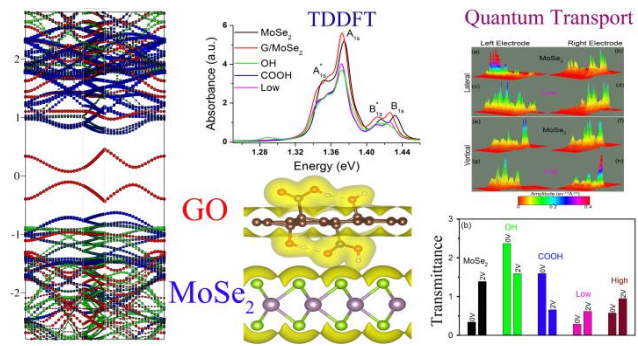

TOC figure only

**Keywords**: Graphene Analogues; Heterostructures; DFT and TDDFT; Quantum Transport; Lateral and Vertical contact; Tunable doping; Ligands

Graphene analogues (GA) belong to the category of two-dimensional (2D) materials where strongly covalent atomic layers devoid of dangling bonds are stacked vertically by van der Waal (vdW) interactions,[1-3] enabling low-energy techniques like mechanical or liquid exfoliation to isolate individual layers.[1-2, 4] Embracing a wide range of materials with different electronic properties like superconductor ($NbSe_2$), semiconductor ($MoS_2$) to insulator (BN), 2D nanosheets of GA systems offer versatile material attributes like high electron mobility[5] and mechanical strength,[6-8] high photoluminescence due to non-centrosymmetry induced direct band gap,[9] systematic valley and spin control by optical helicity, presence of tightly bound trions,[10] excellent channel scalability and high on-off ratio for field-effect devices and so on.[11-12] Variant material properties of these systems highlight the prospect of their utilization for future generation flexible nano and opto-electronics.[1, 6]

Graphene (G), the mostly explored 2D material of last decade,[13] is well-reputed for its extra-ordinary mechanical and transport properties.[13-14] However, materialization of Graphene-era in nano-electronics has stumbled because of its zero band-gap semi-metallic nature and the consequential lack of control on the flow of carriers.[13, 15] Rectification of this drawback has been attempted by various means like structural functionalization or appropriate interfacing.[15-18] Graphene Oxide (GO) is chemically functionalized Graphene, with oxygen containing functional groups randomly decorated on both sides of the G surface,[19] aiming towards modified electronic properties due to the disruption of the $sp^2$ hybridization of G.[20] Transition metal dichalcogenides (TMDC), composed of a sandwich structure containing the transition metal layer in between the two chalchogen layers, are known to overcome the drawbacks of G. Monolayers of TMDC exhibit a direct band gap in the range of 1.5 – 1.8 eV, qualifying their usage in nano-electronic devices.[21-23] The significant deterrent of TMDC systems is their low mobility due to interface-induced scattering effects, where sustained efforts for betterment are pursued by using appropriate substrate and gate material.[24] MoS$_2$ is the mostly studied material of TMDC group because of its intriguing properties.[8, 22-23, 25-26] MoSe$_2$ is the selenium counterpart of MoS$_2$, having a direct monolayer optical gap ~ 1.58 eV at room temperature, which is less than MoS$_2$ (1.9 eV), having an optimal range for optical device utility.[27-28] Bulk MoSe$_2$ exhibits an indirect band gap of ~ 1.1 eV in comparison to ~ 1.2 eV for bulk MoS$_2$.[29] It possesses high trion binding energy (~30 meV) and narrower emission line-width.[30] MoSe$_2$ is also known to be a promising candidate to observe interesting phenomena like exciton condensation[31-32] or Fermi-edge singularity[33] and at the same time it has proved its potential for new generation optoelectronic devices.[34]

An extraordinary privilege of such vdW systems is the uninhibited materialization of combination of disparate systems to obtain tailor made *best of both* properties. There are evidences of heterostructures (HS) resulting from combinatorial lower dimensional systems like 0D-2D, 1D-2D or 2D-2D.[35] Among all these systems 2D-2D HS has established its potential to obtain furtherance in carrier transport, excitonic, optical and vibrational attributes.[35] Such HS paves the way for generation of devices with unique characteristics. In the present work, we intend to overcome the drawbacks in the electrical and electronic properties of MoSe$_2$ by integrating it with a top GO layer. The interlayer coupling of these two layers can be modulated depending on the concentration and type of the functional ligands on GO.[36-38] Most of the monolayer TMDC are heavily *n*-type doped due to defect and charge impurities and possess strong Fermi level pinning near the conduction band.[39-40] Previous works have indicated that MoSe$_2$ FETs are *n*-type, posing good gate control. In spite of having a suitable gap-value, its natural *n*-type propensity restricts its versatility of device usage as junction devices or *p*-type field-effect transistors.[41] However, to develop multipurpose applicability, it is very much required to fabricate both *n* and *p* type devices.

Achieving *p*-type doping in TMDC is challenging due to its atomistic thin dimensions. A limited number of trials to obtain a *p*-type doping in TMDC are available in literature. Laskar *et. al.* reported a *p*-type doping by using Nb as a dopant for large-area few-layer CVD grown $MoS_2$.[42] Area selective plasma immersion by using phosphine plasma has also enabled a *p*-type doping in $MoS_2$.[43] However, in all these processes, the tunable control of doping is lacking. Controlling the doping by growing a bilayer heterostructure is relatively easier to handle. There are several studies in the literature depicting modification of doping, transport and optical properties of the TMDC underneath by placing a top layer of G or GO.[23, 38, 44-46] Depending on the oxygen concentration of the attached GO layer, Musso *et. al.* has demonstrated a lowering of the *p*-type Schottky barrier height for $GO/MoS_2$ system.[38] Nevertheless, these previous studies have not dealt with fine tunability of doping and their impact on non-collinear magnetism, static and time dependent optical properties and realistic-device transport.

In the present investigation we have demonstrated tuning of doping for bilayer HS $GO/MoSe_2$ by varying the concentration and types of functional ligands attached to the GO layer. Effects of functional ligands in modulating the interlayer coupling and thereby obtained impacts on the static and time dependent optical and device transport properties are elaborated. The outcome of these studies suggests the promising potential of the present HS in future opto-electronic device applications.

## Results and Discussions

To study the electronic properties of $GO/MoSe_2$ interface, we have initially constructed a Graphene (G)/$MoSe_2$ interface by joining the two lattices by Co-incidence Site Lattice (CSL) Method, as implemented in the Atomistic Toolkit 15.1 package.[47] In this method, a search was carried out through the grid $m\mathbf{v}_1 + n\mathbf{v}_2$. The vectors $\mathbf{v}_1$ and $\mathbf{v}_2$ are the basis vectors of

MoSe$_2$ lattice. The integer *m* and *n* are chosen in such a way that the supercell of both of the lattices has the lowest mismatch. As a next step to reduce strain, the mutual rotation angle between the two surfaces around the stacking direction is gradually varied in increments of four degrees. In all our calculations, the strain is ~0.66%, when the mutual rotation between a stacking of 5×5×1 G on a 4×4×1 MoSe$_2$ surface about the *z*-axis is ~ 19 degrees. The GO surface was constructed from the 5×5×1 surface of G with varying concentration of hydroxyl (OH) and carboxyl (COOH) functional ligands. We have constructed four different types of GO surfaces, *viz.*, (1) with single OH (G-OH), (2) single COOH (G-COOH), (3) lesser concentrations of both OH and COOH (G-Low) and (4) higher concentrations (G-High) of both ligands. To avoid strain induced buckling of GO layer, the functional groups are placed symmetrically on both sides of the G surface.[37-38] To demonstrate the role of these functional groups in modulating interlayer coupling, we have constructed five HS by following the CSL method to minimize the interfacial strain, *viz.*, (A) G/MoSe$_2$ interface (HS1), (B) G-OH/MoSe$_2$ (HS2), (C) G-COOH/MoSe$_2$ (HS3), (D) G-Low/MoSe$_2$(HS4) and (E) G-High/MoSe$_2$ (HS5). For MoSe$_2$ ML, the most common defect is the Se-vacancy (SeV). The doping trend for all these HS are also investigated in presence of ~ 3% SeV. The ionic positions of the individual interfaces are relaxed to obtain the lowest energy configuration. A vacuum of ~ 15 Å is introduced at top and bottom of the interface to avoid periodic replication from the adjacent supercell. The structures of all GO/MoSe$_2$ interfaces (HS2-HS5) are presented in Fig. S1(a-d).

Our present study on GO/MoSe$_2$ system can be categorized into the following steps: 1) Using first principles technique comprising spin-orbital coupling (SOC), we have investigated the tunable carrier doping and the respective impact on non-collinear magnetic properties of the GO/MoSe$_2$ system; 2) MoSe$_2$ is well-accomplished for its interesting excitonic behaviour. We have investigated the optical properties of the system by static DFT and time-dependent density functional theory (TDDFT) to understand the impact of modulated interlayer coupling on the optical properties of the HS; 3) None of the earlier works have examined an extrapolation of their doping achievement for realistic devices. By using self-consistent DFT-based quantum transport calculations, we have investigated the device made out of the different GO/MoSe$_2$ configurations as a channel material for both edge/lateral and top/vertical contact geometries using Au as contact metal.

**DFT + SOC results:**

The detailed analysis of electronic structure of the GO/MoSe$_2$ heterostructure (HS) is performed by using GGA-PBE + SOC calculations. For such systems lacking inversion symmetry, effect of SOC is significant. We start with a comparison of the GGA and GGA+SOC bandstructure of MoSe$_2$ 4×4×1 monolayer (ML) surface. Fig. 1(a) and (b) depicts the band-structure and atom (APDOS) and orbital projected (OPDOS) density of states of monolayer MoSe$_2$ surface respectively. The peak of the DOS at ~ -1.1 eV corresponds to the $\sigma$-bonded Se-4$p_z$ and Mo-4$d_{xz}$ and 4$d_{yz}$ orbitals. The states at the top of the valence band and bottom of the conduction band in the bonding and antibonding region are mostly constituted of π-bonded Se-4$p_x$, 4$p_y$ and Mo-$d_{xy}$, $d_{x^2-y^2}$ and $d_{3z^2-1}$ orbitals. The GGA band gap obtained for this system is ~ 1.47 eV. For MoSe$_2$, the chalchogen (Se) ligand field-effect is more prominent than its Sulphur counterpart, making the bands wider and thereby reducing the band gap in comparison to MoS$_2$. The band structure along Γ-M-K-Γ high symmetry path clearly shows the direct band-gap at K-point. Figure 1(c) and (d) presents the band-structure and DOS of the same system respectively after incorporating SOC. The band structure reveals a spin-splitting for both valence (VB) and conduction bands (CB). Whereas for the VB splitting, the valley dependence of the spin-splitting is the dominant cause, the CB splitting is a combined effect of the larger negative component of SOC for Se and a smaller positive contribution from Mo-$d$ orbitals.[48] Since incorporation of SOC renders the magnetic moments to be non-collinear, the adjacent DOS of Fig. 1(d) has directionally projected components. For the 2D monolayer system, the *z*-component of the DOS is the most prominent. The calculated band gap of 1.47 eV for non-SOC case reduces to 1.36 eV for SOC.

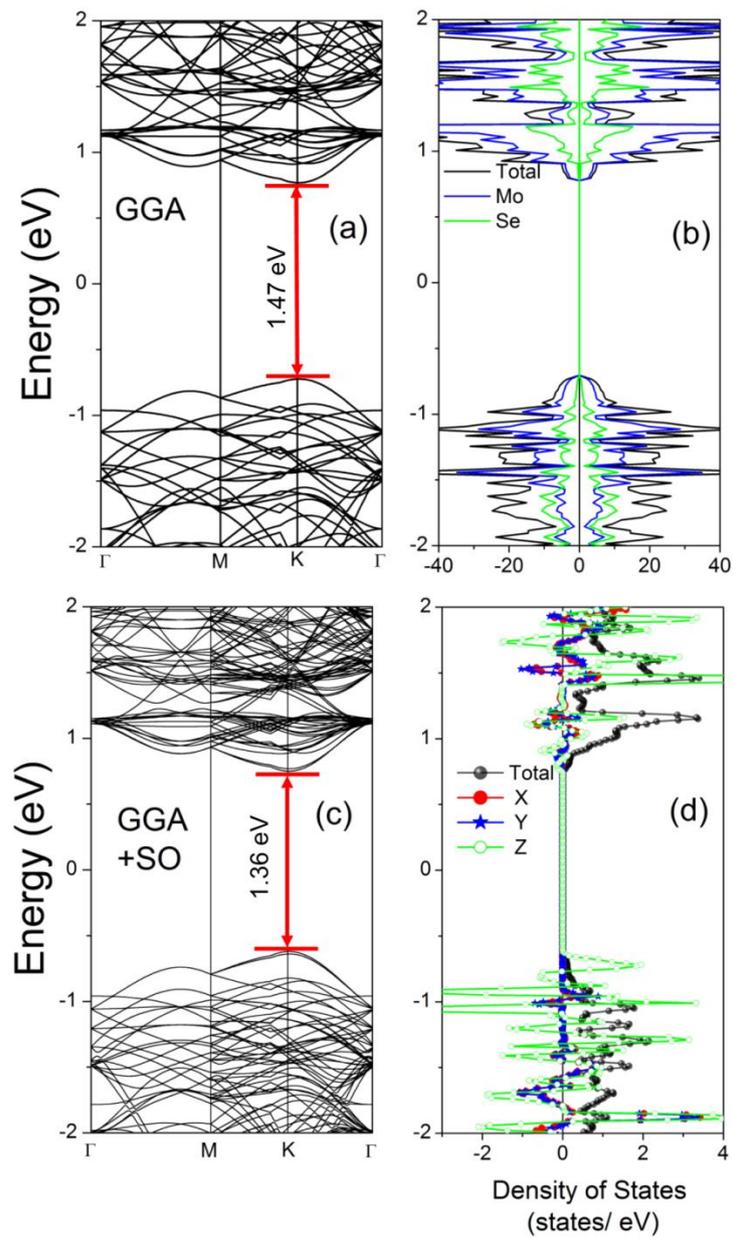

*Figure 1:* (a) GGA-PBE band structure and (b) atom projected density of states (APDOS) of pure MoSe$_2$ structure, (c) Corresponding band structure and (d) DOS of MoSe$_2$ after incorporating SOC.

The ubiquitous effects of SOC for such systems renders its inclusion for all further calculations of HS. For HS1-HS5 and their corresponding SeV cases, as described in the prior section, there is a reduction of ground state energy for GGA+SOC, in comparison to GGA. Presence of ligands at the G-surface can tune the inter-layer charge-transfer and thereby can also control the doping of the underneath $MoSe_2$ layer. Interlayer charge-transfer between the component layers of the HS will be evident from the charge density plots presented in Fig. 2(a) – (d).

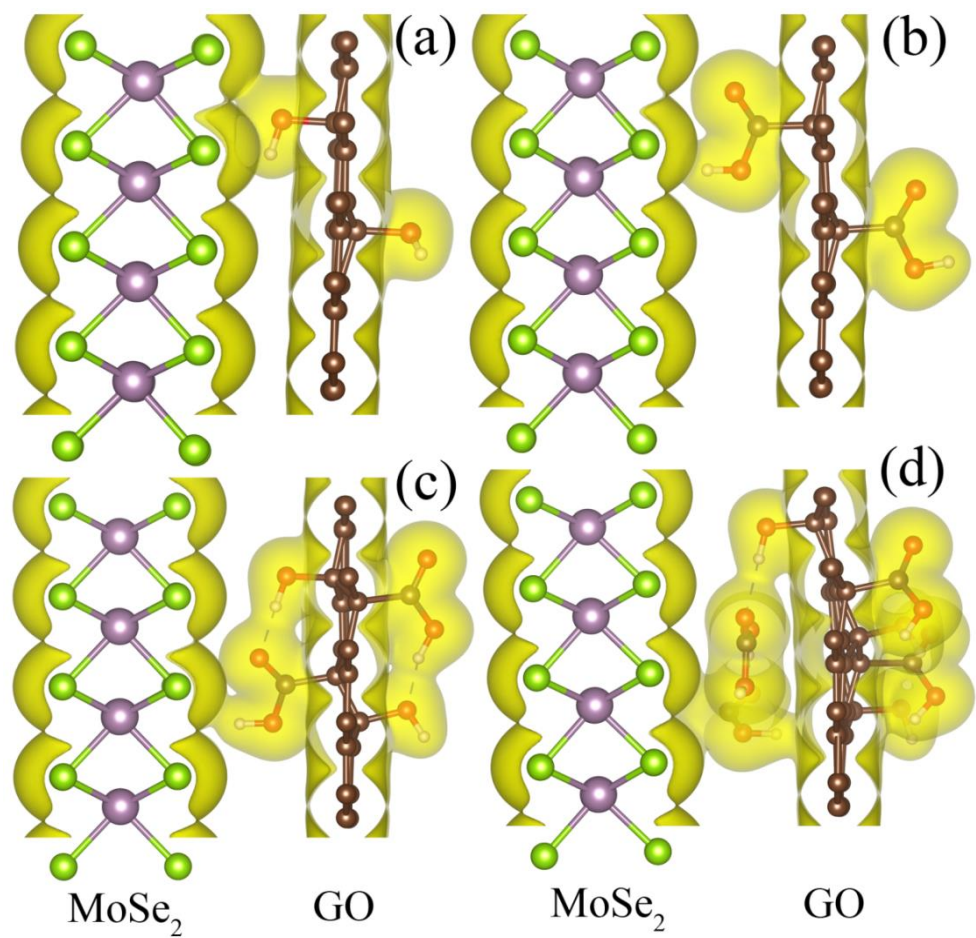

*Figure 2: Converged charge density plots for a) HS2 b) HS3 c) HS4 and d) HS5. Overlapping between the constituent layers indicate the mutual charge transfer. Violet and green represent Mo and Se atoms respectively whereas brown represents carbon atoms. Oxygen atoms are denoted by red colour.*

The type of doping and direction of charge transfer is entirely dependent on the ligand type and concentration. For HS2, where there are single OH group on both sides of G, $MoSe_2$ experiences an *n*-type doping and the corresponding positive shift of the $E_F$ in comparison to the pristine system. Due to lower concentration of ligands, the buckling of the G layer is less. The simultaneous satisfaction of electron affinity of Oxygen of OH ligand from H and C of G-layer promotes the excess electron to flow towards $MoSe_2$. In addition, the interlayer spatial position of H reduces the hybridization of O and Se. Direction of charge transfer reverses in case of HS3, where a more polar functional group, COOH is attached to both sides of G. This results into a larger buckling due to the 3D nature of C-C bond between G and COOH culminating a local conversion of hybridization from $sp^2$ to $sp^3$. Moreover, the higher oxygen content of COOH and its buckling induced vicinity to the $MoSe_2$ ML expedites more hybridization of the *p*-levels of O and Se, leading to the electron transfer from $MoSe_2$ to GO. The $E_F$ of the combined system shifts towards valence band and the system acquires a *p*-type doping. The extent of *p*-type doping increases upto an optimum range with increasing ligand content of GO, keeping an 1:1 ratio of the numbers of OH and COOH ligands. Uncontrolled increase of ligand concentrations leads mostly to two detrimental effects for *p*-type doping: 1) large buckling of the G-layer disturbing the van der Waal stacking, 2) mutual charge transfer between the ligands due to proximity. We have obtained an increase of *p*-type doping upto HS4. For HS5, although the trend of *p*-type doping is retained, the extent of doping is less in comparison to HS4. Presence of SeV is well-known to introduce *n*-type doping. All of these systems show a reduction *p*-type doping in presence of SeV, as described in the supporting information. The relative shift of $E_F$ with respect to the pristine $MoSe_2$ is listed in table 1 for GGA+SOC cases. The trend of doping remains same for both GGA and GGA+SOC cases.

However, the pernicious outcome of increasing ligand concentration is seen to have interesting effects on the SO-coupled non-collinear magnetic properties of the system. From HS2-HS5, with increasing ligand concentration, the $x$, $y$ and $z$ component of the magnetic moment gradually increases, as can be seen from table 2. We have also presented a comparative spin-density plot in Fig. 3, which suggested an increase in spin-density of the systems with increasing ligand concentration for all the three components at both layers. For GO layer, the manifestation of increment of spin-density is more due to its buckling. The same order of magnitudes of magnetic moments for all the three components suggests absence of any magnetic anisotropy.

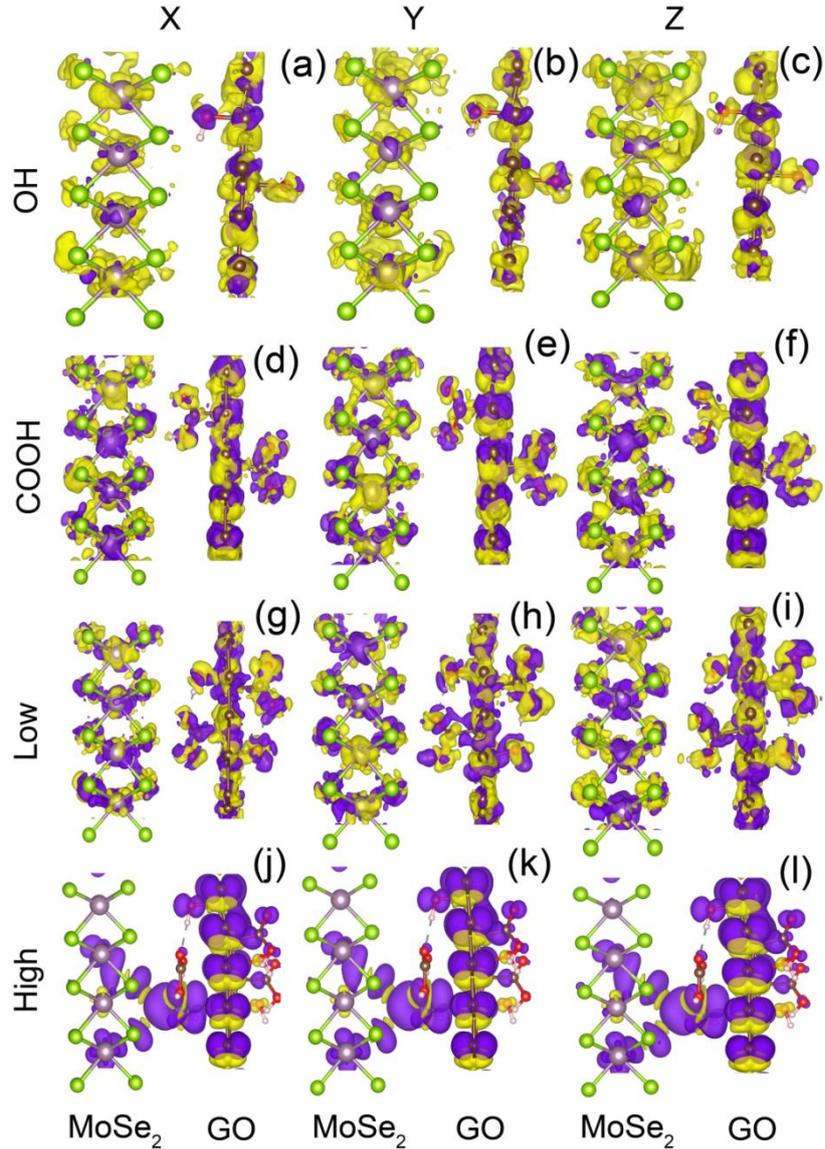

*Figure 3:* Spin Density plots for HS2 for a) x, b) y, c) z-projected components, HS3 d) x, e) y, f) z, HS4 g) x, h) y, i) z and HS5 j) x, k) y, l) z.

To understand the manifestation of ligand concentration on the electronic band-structure of the system, the atomically projected band structures are investigated for three cases: (1) For G/MoSe$_2$ (HS1), (2) G-OH/MoSe$_2$ (HS2) – the *n*-doped case and (3) G-Low/MoSe$_2$ (HS4) – the maximal *p*-doped case, as presented in Fig. 4(a)-(c). In each figure, the fatbands indicating total Mo, Se and C characters are indicated in different colours. For all three HS, the Dirac cones of Graphene bands are split due to vdW interactions with MoSe$_2$. For HS1 (Fig 4(a)), there is a direct band gap between the Dirac cones of G at K-point ~ 0.32 eV.

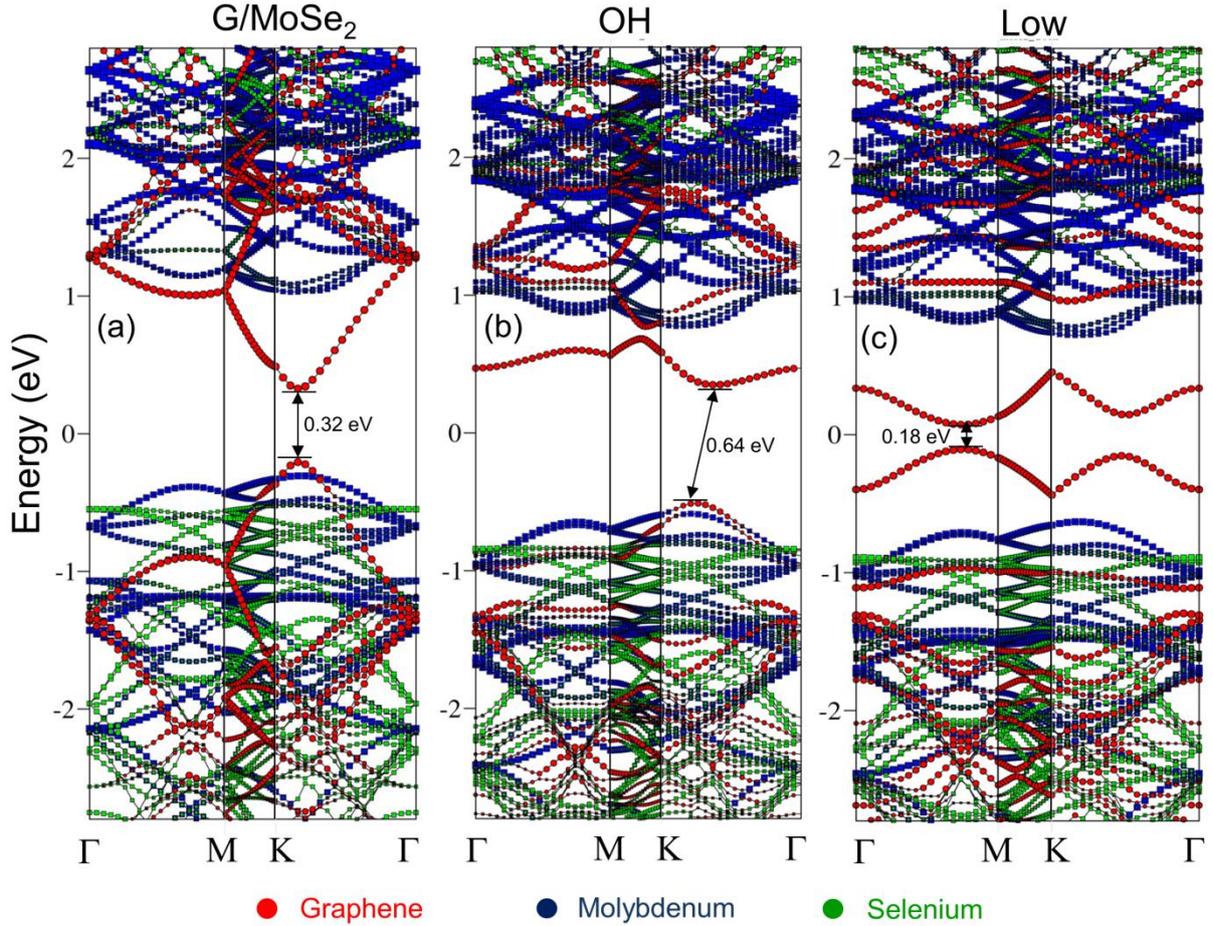

*Figure 4: The atom projected fatbands for a) G/MoSe (HS1)$_2$, b) G-OH/MoSe$_2$ (HS2) and c) G-Low/MoSe$_2$ (HS4).*

Widening of band gap in comparison to the reference 43 is a result of the strain minimizing mutual rotation between the MoSe$_2$ and G surfaces. After attachment of ligands, the Dirac cones of G are gradually detaching from the CB and VB, introducing a modulation of the band-gap magnitude, nature and position. For HS2 (Fig 4(b)), presence of OH on both sides of G, introduces a high degree of hybridization of Mo-$4d_{3z^2-1}$ and C-$2p_z$ at VBM. The unfilled part of the Dirac cone at conduction band is separated down and the band gap becomes indirect at an intermediate position between K to Γ with a value of 0.64 eV. For HS4 (Fig 4(c)), due to symmetric placement of both types of ligands and the increased extent of $sp^3$ hybridization, two symmetrically split C-$2p_z$ Dirac cones are placed around $E_F$. The direct band-gap of 0.18 eV is now situated between Γ to M. A closer look to Fig. 4(a)-(c) also reveals that presence of G and GO on top of MoSe$_2$ leads to a drastic variation of effective mass of carriers, as calculated from the band structure and presented in table 3. In case of HS1, there is a reduction of effective mass to 0.12 from the value of 0.66 for MoSe$_2$, stipulating high carrier mobility for the combined system. Addition of ligands to the G-layer results into flattening of bands and thereby leads to an increase in effective mass. Interestingly, for all three HS, the difference between VBM and CBM of MoSe$_2$ layer remains intact along with their unaltered position in K-space between K to Γ.

Thus, increase of ligand concentration in GO layer is capable of modulating the interlayer coupling and thereby can also modify the carrier-induced electronic properties of GO/MoSe$_2$ HS systems.

**Exciton mechanism for GO/MoSe$_2$ HS from DFT and TDDFT:**

In this section, we have compared the optical absorbance of different HS using static DFT and TDDFT and computed the excitonic positions from the calculations. For 1H-MoSe$_2$, the experimental values of the position of the A and B excitonic transitions are at ~ 1.59 and 1.65 eV respectively.[49]

For static DFT, we have used the Kubo-Greenwood formula to calculate the susceptibility tensor and the real and imaginary part of the dielectric tensor in terms of refractive index ($n$) and the extinction coefficient ($\kappa$) as $\epsilon_r = (n + \iota\kappa)^2$. The absorption coefficient can be derived from the extinction coefficient as: $\alpha_a = 2\frac{\omega}{c}\kappa$.[47] The detailed procedure of calculations is described in the supporting information.

For a one-to-one comparison of the PDOS and the optical absorbance, GGA + SOC PDOS and the corresponding optical spectra are plotted at consecutive columns of Fig. 5 for all five HS along with MoSe$_2$. The first column presents a comparison of the normalised DOS for MoSe$_2$ along with HS1-HS5. In GO, an increase of ligand concentration leads to an increase of hybridization of ligand *p*-orbital with π-orbitals of G, breaking the $sp^2$ bonding chain of G. As a result of this buckling, the intersecting Dirac cones corresponding to the G-band structure open up with appearance of in-gap levels. For HS2, the *n*-type doping results into a GO-induced level with lesser mobile carriers ~ 0.3 eV below the conduction band. For all the three *p*-type doped cases HS3-HS5, there is gradual shrinking of band gap and appearance of acceptor levels near VB, as can be seen from Fig. 5. The position of acceptor level is closest to VB for HS4 (G-Low) case, supporting the fact that this is the most *p*-type doped system. G-High (HS5) possesses a band gap very close to zero.

The initial peak in the absorbance spectrum of $MoSe_2$ (Fig. 5(a)) is at 1.5 eV due to the inter-band transitions from VBM to CBM, corresponding to the optical band gap, which essentially is same as the electronic band-gap for the static DFT calculations, without the correction of exciton binding energy. The transitions are mostly from the Se-$4p_x$, $4p_y$ –hybridized Mo-$4d_{xy}$, $4d_{x^2-y^2}$ and $4d_{3z^2-1}$ filled states at VBM to the unfilled states at CBM having almost the same character.[50-52] This peak at ~ 1.5 eV is present even in the presence of G for HS1 and also with varying concentrations of the ligands from HS2-HS5. Interestingly, for HS4, with optimal ligand concentration to obtain maximal *p*-type doping, this peak is the most prominent. Mutual charge-transfer between G and $MoSe_2$ via the functional ligands, albeit appearance of mid-gap states, does not modify the difference of VBM and CBM of original $MoSe_2$ layer, as also seen in Fig. 4. The origin of peak of optical absorbance at ~ 1.5 eV is basically due to the transitions from VBM to CBM.

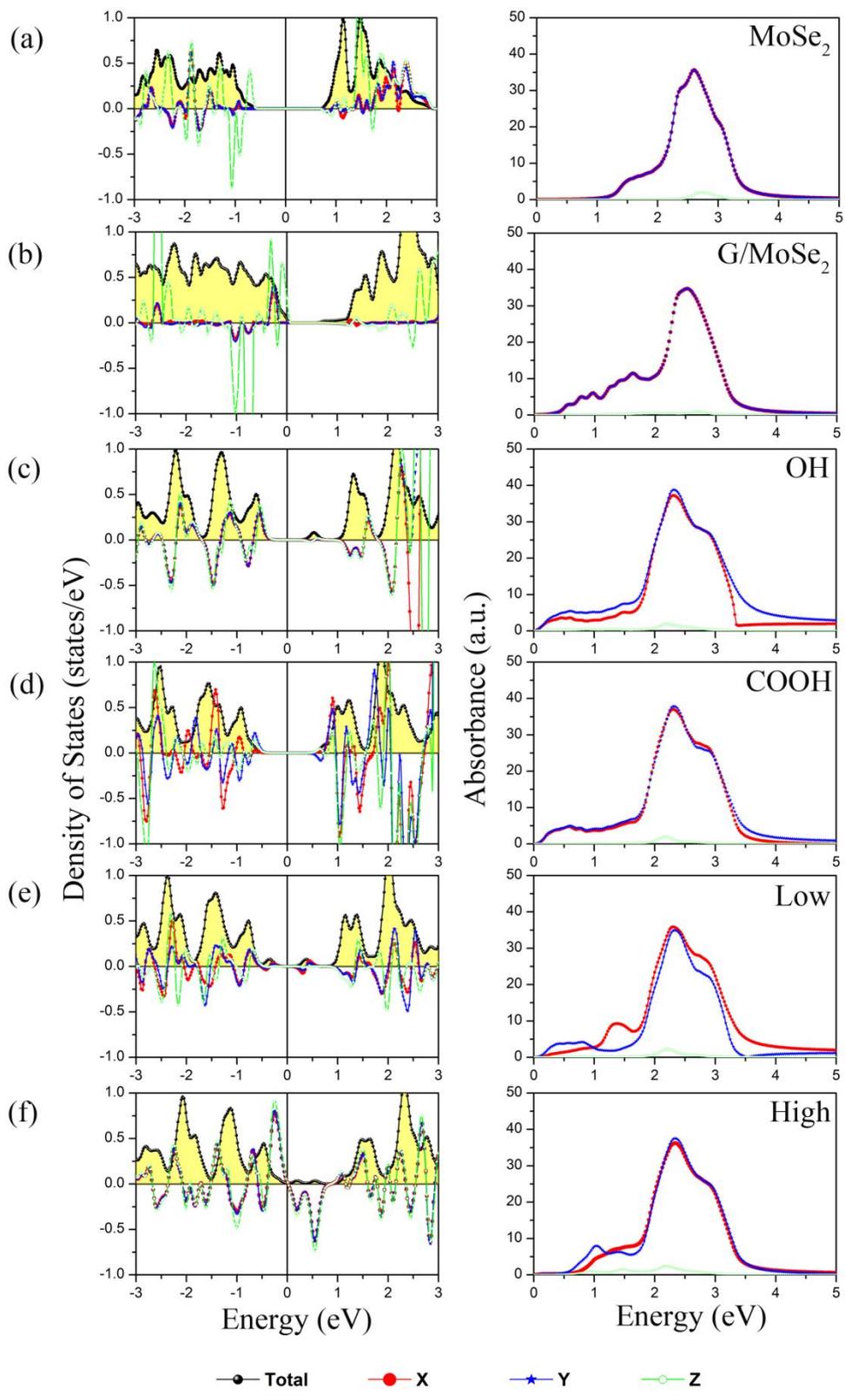

*Figure* **5:** *The SO coupled PDOS and corresponding absorption coefficients of GO/MoSe$_2$ with different ligand concentration, a) MoSe$_2$, b) HS1, c) HS2, d) HS3, e)HS4 and f) HS5. Red, blue and green colour represents x, y, z directions respectively.*

The highest peak in absorbance occurs at ~ 2.5 eV, corresponding to the transition from the strong bonding levels of MoSe$_2$, consisting of Se-4$p_z$ hybridized Mo-4$d_{xy}$, 4$d_{x2-y2}$ and 4$d_{3z2-1}$ levels to the unfilled Se-4$p_x$, 4$p_y$ - hybridized levels of same orbital Mo-character.[53] From the PDOS plot 5(a), it will be evident that the contribution to the z-component is more at the bonding regime. For HS1, due to the presence of G, the bandgap is reduced, the outcome of which is revealed in the absorbance spectrum with the appearance of additional peaks before 1.5 eV. There is, however, no change in the highest peak position of the absorbance for HS1 (~2.5 eV), implying that the presence of an overlayer G does not have much impact on bonding orbitals. With the variation of ligand concentrations from HS2-HS5, the highest peak shifts at a slightly lower energy of ~ 2.3 eV, as can be seen from Fig. 5(c) to (f).

To obtain an idea about the excitonic mechanism in such HS, we have also investigated the time dependent optical properties of these systems. For this purpose, we have used time-dependent density functional theory (TDDFT) approach, as implemented in ELK code.[54] The computational methodology is described in computational section. TDDFT is an extension of density functional theory in time domain, where time dependence is incorporated in the approximation of the exchange-correlation kernel (XCK) from the explicit time dependence of the xc potential and electron-density.[55]

We have used the bootstrap kernel to investigate the excited state behaviour and formation of bound excitons, as described in the supporting information. The TDDFT kernel, possessing correct long range 1/$q^2$ behaviour in the long-wavelength limit, is capable of reproducing the formation of bound excitons with an optimal computational cost.[56-58]

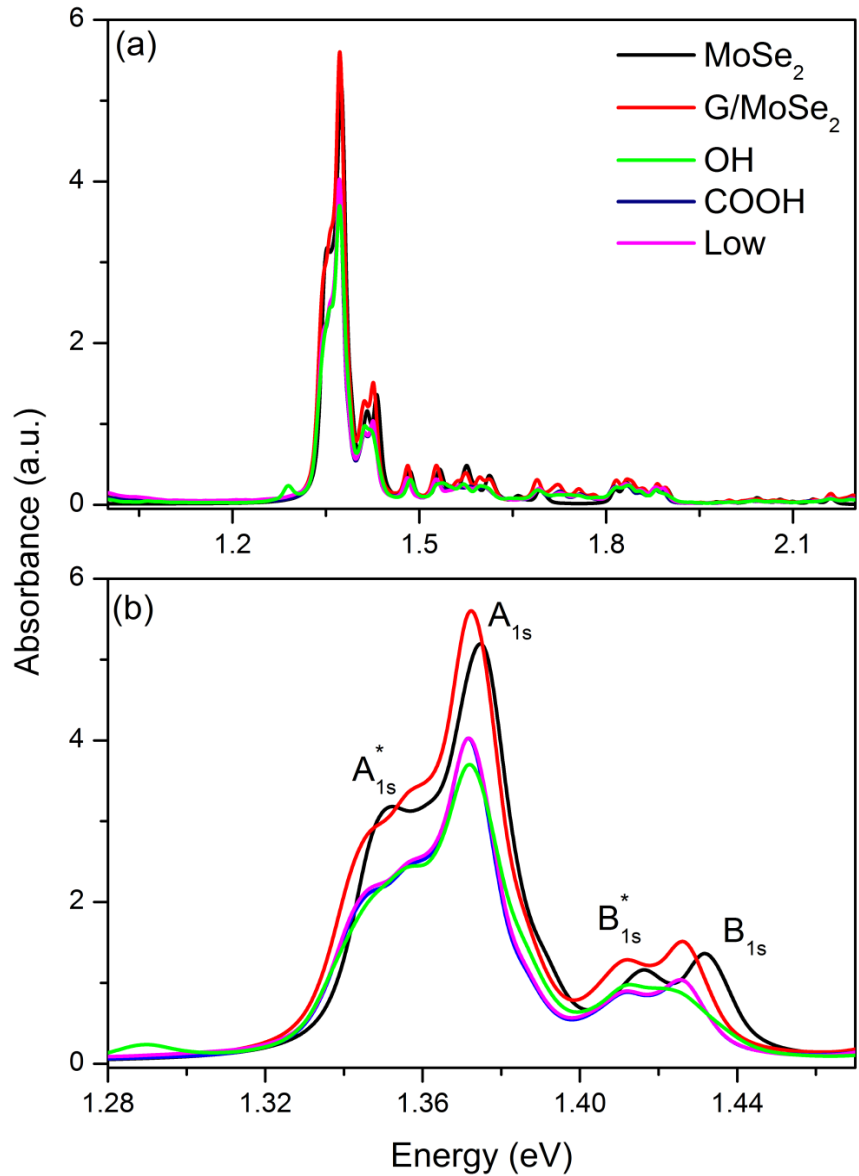

*Figure 6:* *The absorption coefficient of GO/MoSe$_2$ with different ligand concentration, a) MoSe$_2$, b) OH, c) COOH and d) Low, calculated using TDDFT method.*

Figure 6 represents a comparison of the absorption coefficient calculated using TDDFT for different GO/MoSe$_2$ HS. Since performing such calculation on the earlier HS is computationally expensive, we have constructed a smaller HS of MoSe$_2$ and GO containing different concentrations of functional ligands. Strain minimization and ionic relaxations are performed on these systems to obtain the ground state configurations. Calculations are carried out for MoSe$_2$ monolayer, G/MoSe$_2$ interface (HS1), G-OH/MoSe$_2$ (HS2), G-COOH/MoSe$_2$ (HS3) and for (E) G-Low/MoSe$_2$ (HS4). Achieving a high concentration of ligands is not plausible for this smaller structure. For monolayer MoSe$_2$, the absorbance depicts two main excitonic peaks and their corresponding low-energy satellites, which may be due to bound charged excitons like trions.[59] The presence of positively and negatively charged quasiparticles are well-known for such systems.[10] Presence of A$_{1s}$ and B$_{1s}$ exciton positions are calculated to be at 1.38 and 1.43 eV respectively (Fig. 6b). The corresponding experimental values are 1.59 and 1.65 eV.[49] The peak positions are slightly underestimated in TDDFT, which may be due to the approximations used for XCK. The approximation used for exchange correlation (XC) functional does not seem to have much impact. Repetitions of same calculations with GGA functional have not changed the exciton peak positions. The corresponding charged exciton peaks A$_{1s}^*$ and B$_{1s}^*$ are at 1.35 and 1.41 eV respectively, as can be seen from Fig 6(b). The relative peak shift of trion positions with respect to the primary 1$s$ exciton peaks are around 30 and 20 meV respectively for A and B excitons. In comparison to MoSe$_2$, its interface with G (HS1) has a red-shifted A$_{1s}$ peak, whereas the A$_{1s}^*$ peak is slightly quenched and has a blue shift, implying a lesser bound weaker trion for the interface. This supports the weak $p$-type doping and thereby lacking negatively charged carriers obtained for HS1 with respect to the pristine system. Interestingly, addition of functional ligands does not introduce any energy shift for the A$_{1s}$ or A$_{1s}^*$ peaks. With more $p$-type doping in the system for HS3 or HS4, the intensity of trion peak is quenched. On the contrary, for HS2 with OH introduced $n$-type doping, A$_{1s}^*$ peak is stronger because of more $n$-type carriers. The unchanged peak position for A$_{1s}$ supports the well-known phenomena of band-gap renormalization.[60] The change in the electron-hole interaction for 1$s$ exciton obtained after modifying the electrostatic screening by introduction of different functional ligands at the interface is nullified by the modification of electron-electron interaction and thereby resulting change in the quasiparticle band-gap.[60] For B$_{1s}$ and B$_{1s}^*$ the absorption spectrum of HS1 is red-shifted with respect to monolayer MoSe$_2$. The intensity of all B exciton peaks are less than A exciton. However, for both $B_{1s}$ and $B_{1s}^*$, there is a small modification of peak positions from HS1-HS4 (Fig. 6b). For $B_{1s}^*$, a similar trend of intensity quenching with increasing *p*-type doping is also observed. The higher excitonic peaks are all having lesser intensity and also there are variations of peak positions from HS1 to HS4.

Therefore, presence of functional ligands and the resulting modified interlayer coupling and dielectric screening has impacts for higher excited state absorptions. For low energy excitons, band-gap renormalization effect does not allow any shift of exciton peak positions.

**Quantum transport properties for vertical and lateral contact devices:**

Although both experimental and theoretical accomplishment of anticipated doping pattern by forming the 2D-2D HS is feasible, behaviour of the HS after placement of electrical contacts may undergo drastic changes. The contact geometry and electronic properties of interfaces of metal contacts with the HS of two semiconducting systems have their individual roles to play. We intend to explore the transport behaviour of the HS for lateral and vertical placement of contacts with a commonly used metal like gold (Au). A model device configuration having two electrodes, electrode extension and a central region is shown in Scheme 1. The geometry of the lateral and vertical contacts for real system will be evident from Scheme 1b and 1c respectively. The device configurations are constructed by using the ATK 15.1 package, as described in the computational methods section.

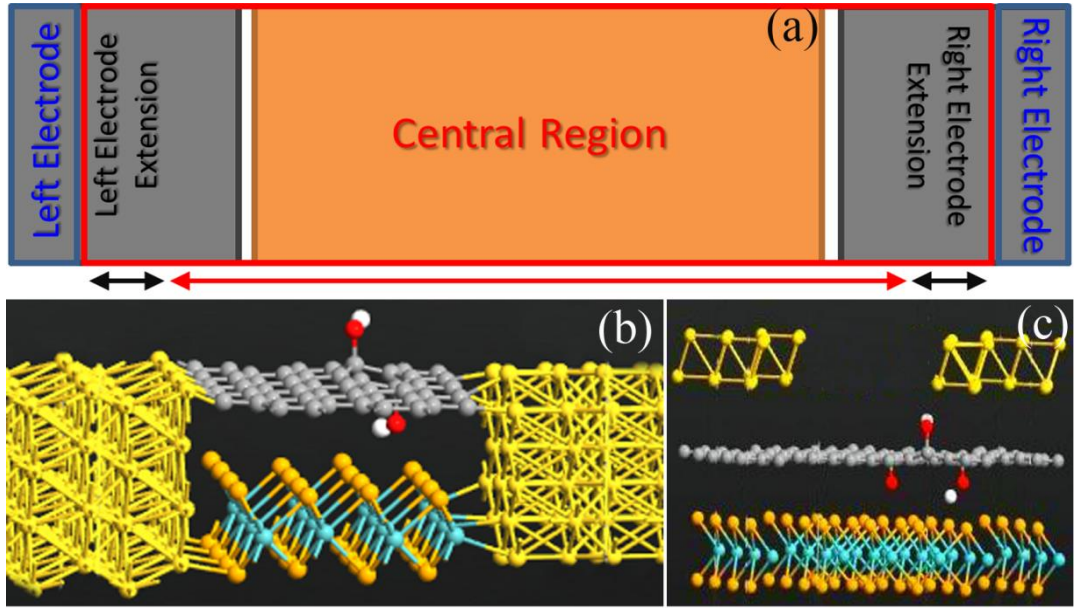

**Scheme 1:** *(a) A model device configuration having two electrodes, electrode extension and a central region, (b) model lateral contact and (c) model vertical contact.*

For MoSe$_2$ ML and for all HS starting from HS1 to HS5, we construct the lateral and vertical interfaces with bilayer Au[111] surface. For vertical contact, the interface of all HS with Au[111] surface was constructed after rotation induced strain minimization procedure to restrict the strain to less than 1%, as described in the methods section. These interfacial systems are relaxed using the conjugate gradient algorithm of VASP.[61] The LPDOS and APDOS for the lateral and vertical systems are plotted in Fig. 7 for all four HS along with the pristine system. A close comparison of the Au lateral and vertical layers on ML MoSe$_2$ unveils the inter-layer charge transfer process and the resulting alteration of doping trend. For lateral contacts, the edge Mo and Se atoms covalently bond with Au. The Mo and Se atoms inside channel, on the other hand, have no hybridization with Au. Au 6$s$ delocalized electrons are transferred to both Mo-4$d$ and Se-4$p$ states leading to almost same weightage of highly hybridized Mo and Se-induced deep bonding levels. The antibonding and bonding states in proximity of E$_F$ have dominance of Mo-4$d$ states mostly because of channel. The situation alters for vertical interface, where both of the bonding and antibonding DOS are dominated by Se-4*p* levels (Fig. 7). In this case, the charge transfer from Au-6*s* to Mo-4*d* occurs via Se-4*p* leading to filled Mo-4*d* states below -1 eV and above 1 eV. Near $E_F$ states are mostly populated by Au and Se hybridized states. The *p*-type schottky barrier (SB) is shorter than *n*-type ones for lateral case, whereas the *n*-type one is shorter for vertical interface.

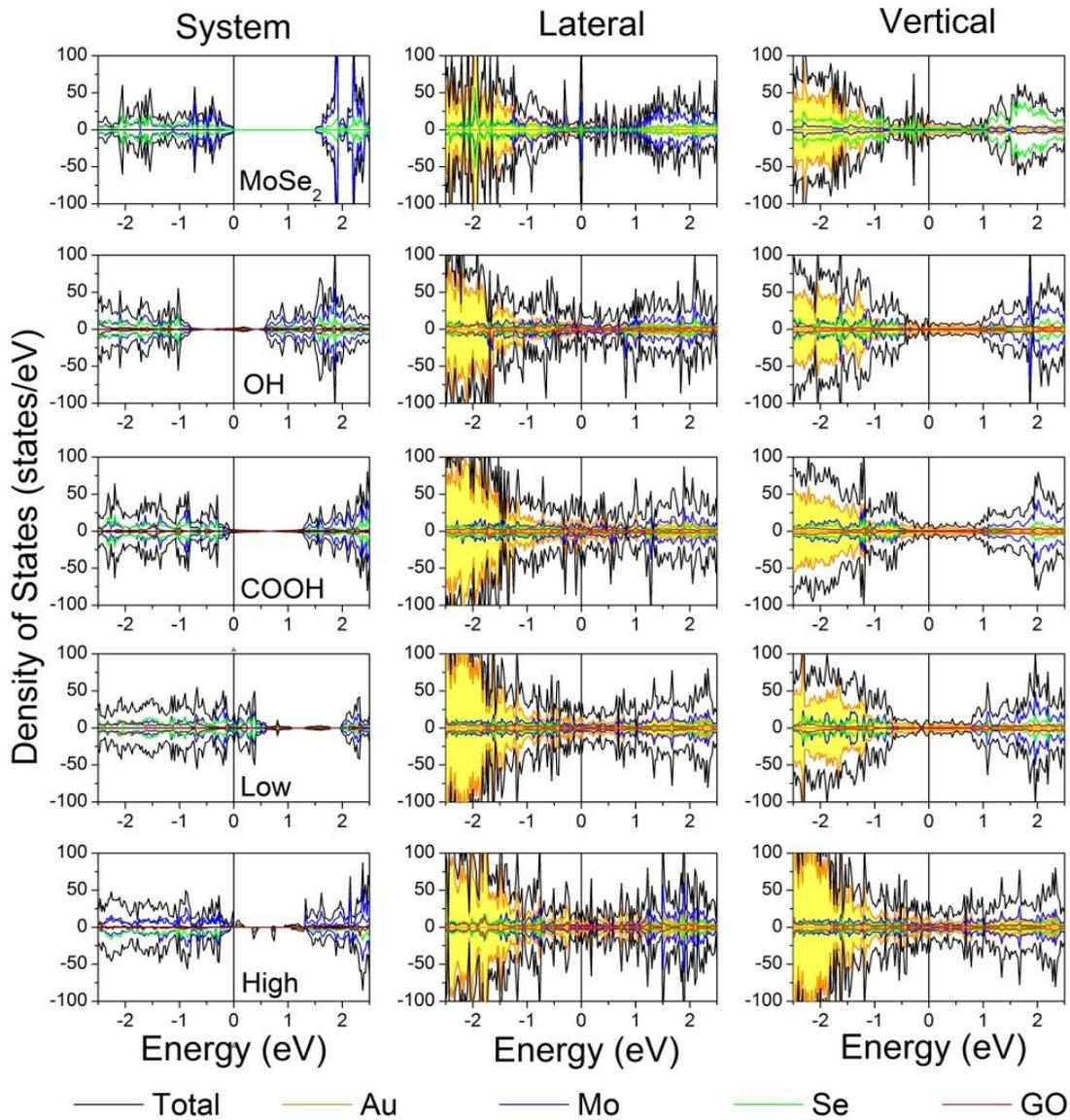

***Figure 7:*** *The APDOS of the pristine system and different GO/ MoSe₂ HS and the corresponding DOS for lateral and vertical interfaces with Au.*

In presence of GO, there is a little modification in case of lateral interface. The transferred charges of Au-6$s$ are now shared between the edge atoms of MoSe$_2$ and GO resulting into complete delocalization of Mo, Se and GO levels and introducing metallic behaviour in the system. On the contrary, for vertical interfaces, the intermediate GO layer reduces hybridization of Se and Au levels and results into disappearance of highly hybridized peaks of Au and Se between 0 to -1 eV as present for ML-MoSe$_2$ system. As a result, near E$_F$, there are delocalized levels of GO and Au. For all four HS (HS2-HS4), the hybridization of Mo and Se levels are regained for vertical interfaces, where both bonding and antibonding levels are dominated by Mo-4$d$ states. In presence of GO, the $p$-type SB is smaller than the $n$-type SB for the underneath MoSe$_2$ layer. Presence of functional ligands on top surface of GO helps the charge transfer between GO and Au and the ligands beneath GO leads to charge transfer between MoSe$_2$ and GO. Therefore, the extent of $p$-type doping achieved because of charge transfer from MoSe$_2$ to GO is reduced in presence of Au interface.

For lateral contacts, the side interfaces are constructed from gold with a strain on gold layer less than 5% on both sides. This system is relaxed and then the device is constructed after extending the right and left electrodes retaining the periodicity. For vertical contacts, the relaxed structure of the interface is considered as the channel and after swapping the $b$ and $c$ axis, both sides of this channel are periodically extended to construct the left and right electrodes. The top Au bilayer on the channel is removed after keeping the Au-layer only at electrodes as top contact. Consequently, for both types of devices, a self-consistent quantum transport calculations are carried out to have an idea of the transmission coefficient and the I-V characteristics as a function of source to drain bias.

The transmission coefficients at zero bias perpendicular to the transmission direction within the irreducible Brillouin zone (IBZ) are calculated for both the device geometries by using the relation:

$$T^{\|}(E) = Tr\left[\Gamma_L^{\|}(E)G_{\|}(E)\Gamma_R^{\|}(E)G_{\|}^{\dagger}(E)\right]$$

where, $G_{\|}$ is the retarded Green's function and $\Gamma_{L/R}$ is the level broadening with respect to the corresponding self-energies of the respective electrodes. This function, while averaged over the *k*-point mesh in the IBZ, results the transmission coefficient.

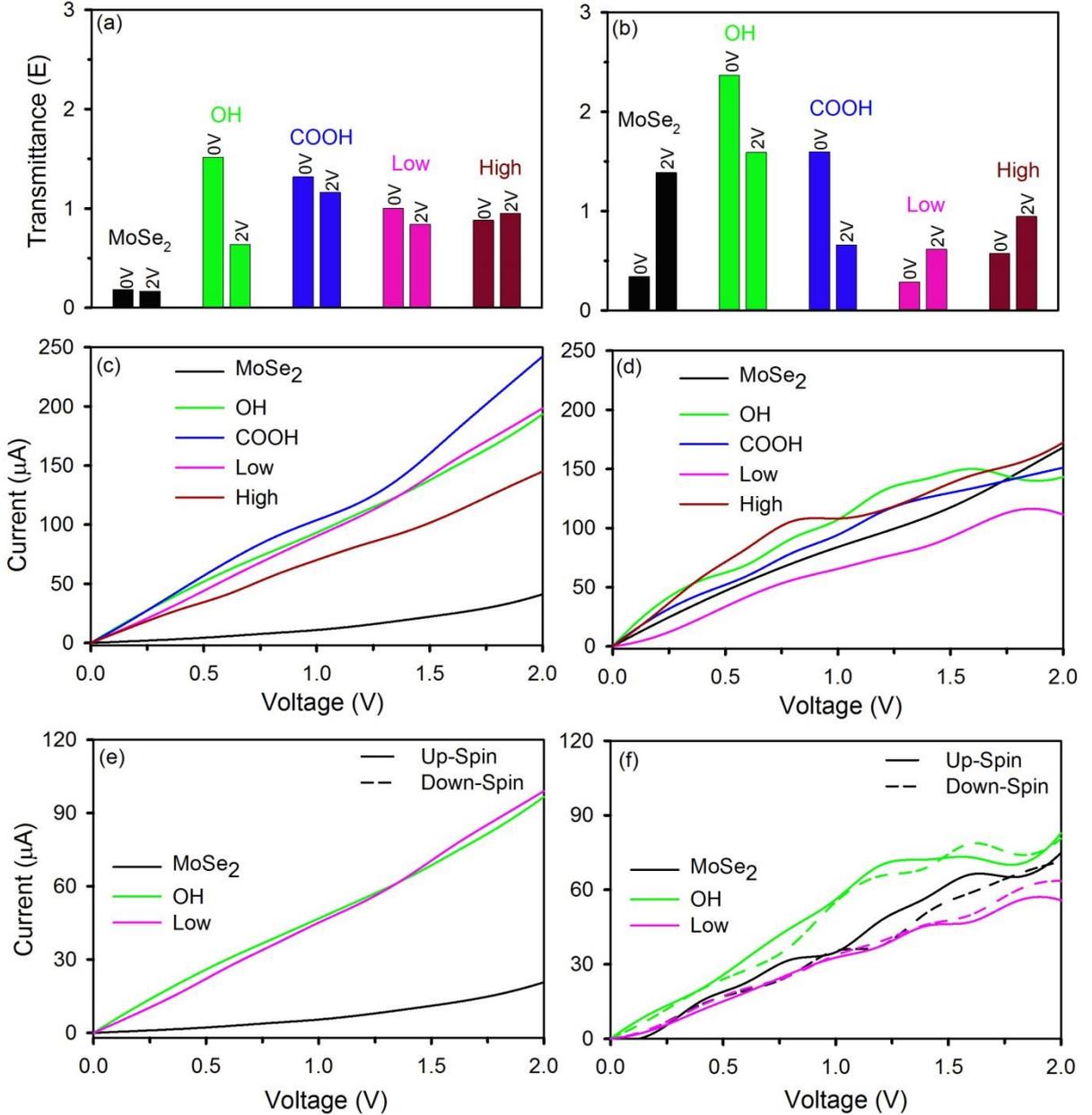

***Figure 8:*** *$\Gamma$ –point transmission of HS and pristine system at zero and 2V bias of a) lateral and b) vertical contact. The I-V characteristics of HS and pure system with c) lateral and d) vertical contact configurations. The spin polarized I-V characteristic of HS and pure system with e) lateral and f) vertical contact configurations.*

Figure 8(a) – (f) depicts a comparison of the transport characteristics of lateral and vertical contact devices for a non-spin polarized as well as spin polarized quantum transport calculations. A comparison of the transmission coefficients for minimum (0 V) and maximum (2 V) bias voltages for both type of contacts reveals that the achieved nature of doping after combining GO with $MoSe_2$ does not extrapolate for devices. For 2D TMDC systems, in general, lateral contacts are known to possess an absence of Schottky barriers (SB) and lower tunnel barriers because of the orbital overlap due to direct chemical bonding and are thereby an efficient low-resistive electron injector. The vertical contacts, although have larger contact area, are more resistive due to the absence of vertical chemical bonding for closed covalent systems like $MoSe_2$, where there are no dangling bonds in absence of vacancies. However, the systems studied in the present work are different from common TMDC systems in many respects. As can be seen from 8(a) and (b), with lateral Au contacts, all of the HS from HS2-HS5 are lower resistive and possess higher transmission coefficients for both bias voltages in comparison to ML-$MoSe_2$, confirming an *n*-type doping with increasing ligand concentration. For vertical contacts, on the contrary, the transmission coefficients follow the expected *p*-type doping trend. The overall current value is lower than the lateral ones, as can be seen from the I-V characteristics plots for both non-spin-polarized and spin-polarized transports in Fig 8(c) – (f). It may also be mentioned in passing that the dependence of transport properties on spin-polarization is manifested only in case of vertical contacts. Transports for lateral contacts have no spin-polarization dependence. With an optimal *p*-type doping for G-Low/$MoSe_2$ (HS4) system, the corresponding vertical contact device is the most resistive one with the current lower than the pristine system for both bias values. Therefore, for GO/$MoSe_2$ HS with optimal *p*-type doping, vertical contact devices are seen to be more effective to retain the achieved doping type even in the presence of metal contacts.

For the present type of HS, the functional ligands on GO modulates the interlayer coupling and thereby control the doping of the underneath MoSe$_2$ ML. The inter-layer vertical charge transfer responsible for controlling doping lowers the height of the SB at the metal-semiconductor junction and thus proves to be more effective to retain the obtained doping. Lateral contacts, on the other hand, having chemical bonding and orbital overlap for both GO and MoSe$_2$ layer, are disruptive for the doping obtained via inter-layer charge transfer.

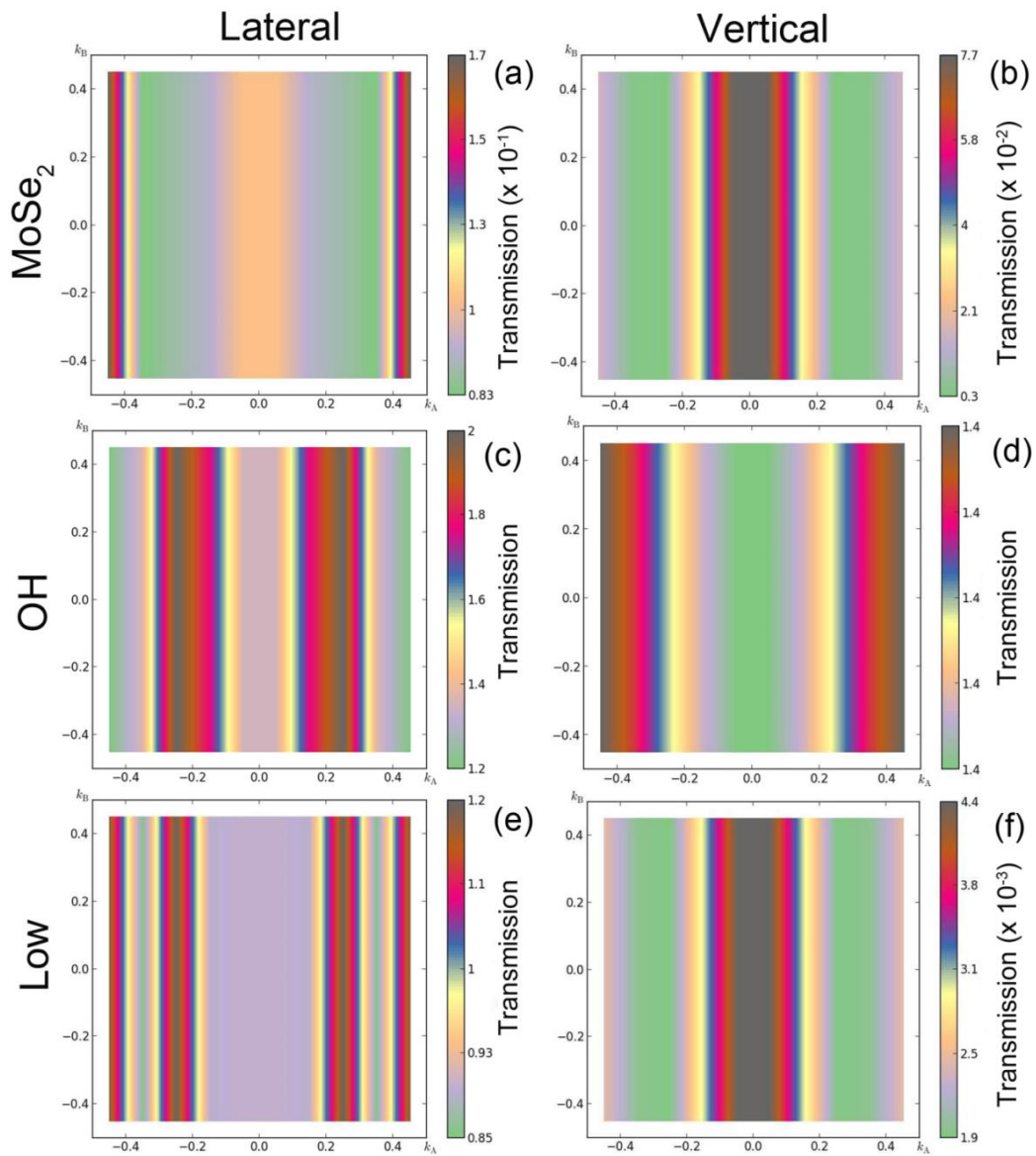

*Figure 9:* $\Gamma$ *–point interpolated contour plot of the transmission coefficients vs. reciprocal vectors at zero bias of pristine MoSe$_2$ a) lateral contact and b) vertical contact, MoSe$_2$ HS with OH (HS2) c) lateral and d) vertical contact and MoSe$_2$ HS with Low (HS4) e) lateral and f) vertical contact.*

Figure 9 presents the Γ-point interpolated contour plots at the ($k_x$, $k_y$) plane perpendicular to the direction of carrier transport for the transmission coefficients corresponding to the pristine system (a and b), HS2 (with G-OH, c and d) and HS4 (with G-Low, e and f). For lateral contacts in pristine system, the central region around Γ is broad and having an overall better transmission than the vertical system. The edges are highly transmitting than the central zone. For vertical contacts, on the contrary, transmission is more for central region than the edges. For HS2 lateral contacts (with OH), the edges are having lower transmission than intermediate regions containing the functional ligands (9c). For vertical contacts, the variation of transmission is very less, as can be seen from the scale, although having higher values at the edges (9d). For HS4 lateral contacts, the edges and ligand regions are more transmitting. For vertical contact (9f), there is a significant reduction of transmission in comparison with HS2 (9d), as evident from the scale.

Figure 10 represents a comparison of the transmission eigenstates for pristine system and HS4 (with G-Low). The average amplitudes of the eigenstates over the channel area are higher for lateral contacts while compared to vertical contacts. Compared to $MoSe_2$, average amplitudes are lower in case of HS4 for vertical contacts for both left and right electrode transmission, implying an altogether lower transmission for HS4. For lateral contacts, on the other hand, the average amplitudes for HS4 are higher than $MoSe_2$ for both electrode transmissions.

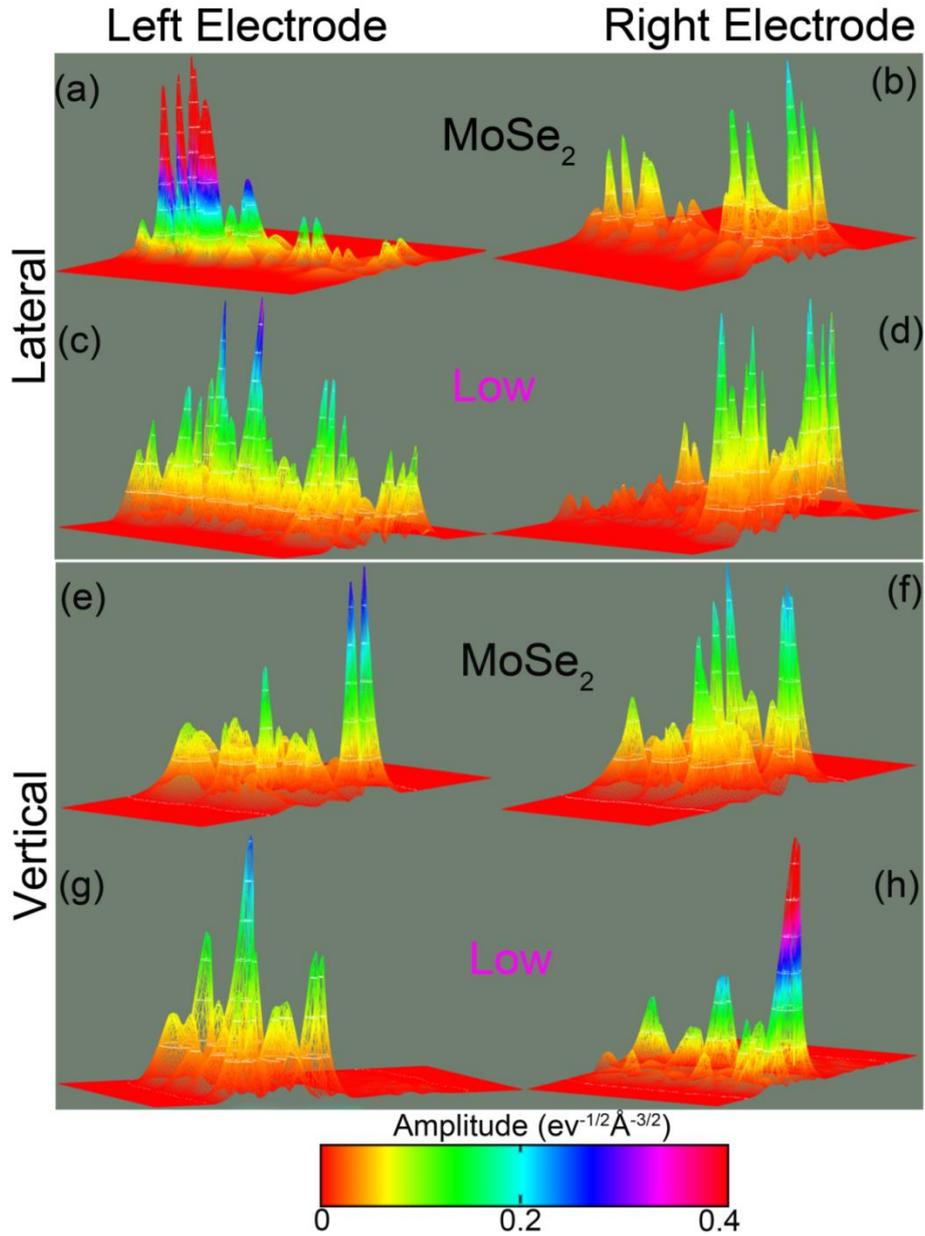

**Figure 10:** *Transmission eigenstates of pristine MoSe$_2$ and Low (HS4) systems depicting transmission amplitudes from both electrodes corresponding to Gamma point transmission, a) Lateral Left (onlyMoSe$_2$), b) Lateral right (only MoSe$_2$), c) Lateral Left (Low), d)Lateral right (Low) e) Vertical Left (only MoSe$_2$), f) Vertical right (only MoSe$_2$), g) Vertical Left (Low), h) Vertical right (Low).*

In summary, we have investigated the fundamental electronic properties of GO/MoSe$_2$ HS and its excitonic and transport properties as a function of the interlayer coupling to analyse the potential of the system for nano-electronic and opto-electronic device application.

## Conclusion

We have investigated the role of functional ligands in modulating the interlayer coupling and the consequential control on the type of doping for various HS of GO/MoSe$_2$ accomplished after varying the concentration and type of ligands. A comprehensive observation of time independent and dependent optical properties has enabled us to identify the interconnection of the underlying modification of electronic structure upon variation of interlayer interactions and the exciton dynamics. We have also studied the realistic device transports for the system with both lateral and vertical contacts with Au. For such systems, where the carrier density modifications are obtained by vertical charge transfer, vertical/top contacts are seen to be most useful to retain the obtained advantage of carrier density. These observations are believed to provide an efficient insight of the electronic structure of the system and thus utilization of these systems for opto-electronic and nano-electronic device applications.

## Computational Details:

We have combined appropriate calculation procedures at different steps of investigations. For time-independent density functional investigations, we have used spin-polarized plane-wave pseudopotential calculations with norm-conserving projector augmented wave (PAW) pseudopotentials as implemented in Vienna Ab-initio Simulation Package (VASP). The valence levels for Mo consist of 4$p$, 5$s$ and 4$d$ orbitals and those for Se constitutes 4$s$ and 4$p$ orbitals. The exchange-correlation interactions are treated with generalized gradient approximation (GGA) with Perdew-Burke-Ernzerhoff (PBE) functionals. The energetics and the magnetic properties are studied after taking into account the spin-orbit coupling (SOC). Interface-induced dipolar interactions are treated by taking care of van der Waals corrections after including a semi-empirical dispersion potential to the DFT energy functional as according to the Grimme DFT-D2 method.[62] The cut-off energy for the plane wave expansion is set as 500 eV and a Monkhorst-Pack grid 5×5×3 is used for Brillouin zone sampling for all calculations. The ionic positions and the lattice parameters are relaxed by using the conjugate gradient algorithm until the Hellmann-Feynman force on each ion is less that 0.01 eV.

The static optical properties and the DFT quantum transport calculations are performed by using the Atomistic Toolkit 15.1 package, with the GGA-PBE variant of exchange correlation.[47, 63] Double-zeta polarized basis sets are used for expansion of electronic density. Contribution of non-local dispersive forces like van der Waal interactions are taken into account by following DFT-D3 method. For calculation of optical properties, a finer *k*-point mesh of 11×11×11 was used with a broadening of 0.1 eV after including SOC. For quantum transport calculations, a DFT coupled Non-equilibrium Green's Function (NEGF) methodology is used. We have constructed two types of contact geometry for the devices, *viz.*, lateral and vertical contacts with Gold (Au) atoms keeping the same channel length. The direction of placement of the electrodes is taken along *c*-axis. For lateral contact, the interface is oriented to get the lateral electrode placement along *c*-axis. The Monkhorst-Pack *k*-point grid for the lateral and vertical contact devices is taken as 5×5×50. The temperature is kept at 300 K and the real-space density mesh cutoff is 200 Hartree with the maximum force of 0.01 eV/A used for geometry optimization. At the boundary of the electrode and the drain-channel, Dirichlet boundary conditions are used to provide charge-neutrality between the source and the drain. Both spin-polarized and non-polarized transports are studied and the corresponding transmission coefficients are calculated by averaging over a *k*-mesh of 10×10 in a direction perpendicular to the current transmission.

The time dependent optical properties are calculated by using the all electron full-potential linearized augmented plane wave approach including local orbitals (FP-LAPW + lo) within the framework of DFT as implemented in the ELK4.3.6 code.[54] We have treated the exchange-correlation potentials with local density approximation (LDA) with Perdew-Wang / Ceperley Alder functional. The basis functions are expanded as linear combinations of spherical harmonics and plane waves within the non-overlapping muffin-tin spheres and at the interstitial regions respectively. The muffin-tin radii of Mo, Se, O, C and H are taken as 2.47, 2.28, 0.99, 1.25 and 0.78 a.u. respectively. The interstitial plane wave vector cut-off $K_{max}$ is chosen such that $R_{mt}K_{max}$ equals 7 for all the calculations, where $R_{mt}$ is the smallest of all atomic sphere radii. This convergence parameter controls the size of the basis. The convergence criteria for total energy and RMS change in Kohn-Sham potential are kept as 0.0001 and 0.000001 eV respectively. The valence wave functions inside the spheres are expanded upto $l_{max}=10$ and the charge density was Fourier expanded upto $G_{max} = 12$.

## ACKNOWLEDGEMENT

TKM wishes to acknowledge the support of DST India for INSPIRE Research Fellowship and SNBNCBS for funding. We also thank DAE (India) for financial grant 2013/37P/73/BRNS. DK would like to acknowledge BARC ANUPAM supercomputing facility for computational resources.

## Associated Content

### Supporting Information
The Supporting Information is available free of charge on the ACS Publications website.

 The supplementary information contains the model geometric structure of GO/MoSe$_2$ HS system. Details of static DFT and TDDFT methodology to calculate the optical properties are described. The Fermi shift and magnetic moment of the HS with SeV has been listed in the supplementary article.

**Table 1:** Shift of $E_F$ with respect to pristine $MoSe_2$

| System | Shift of $E_F$ (eV) | Type of doping |
|---|---|---|
| OH | +0.399 | $n$-type |
| COOH | -0.721 | $p$-type |
| Low | -1.182 | $p$-type |
| High | -0.994 | $p$-type |

**Table 2:** Magnetic moment table of the system with and without spin orbit coupling

| System | Magnetic Moment (SOC) | | | Mag. (Non-SOC) |
|---|---|---|---|---|
| | X Comp. | Y Comp. | Z Comp. | |
| OH | -0.0007 | -0.0007 | -0.0008 | -0.0001 |
| COOH | 0.0003 | 0.0003 | 0.0004 | 0.0003 |
| Low | 0.0005 | 0.0005 | 0.0006 | 0.0007 |
| High | 1.0755 | 1.0719 | 1.3029 | 2.002 |

**Table 3:** Effective mass of the systems calculated from band structure

| System | Effective Mass |
|---|---|
| $MoSe_2$ | 0.66 |
| $G/MoSe_2$ | 0.12 |
| OH | 0.55 |
| Low | 1.05 |